\documentclass[aps,prl,twocolumn,groupedaddress,showpacs,showkeys]{revtex4}
\usepackage{amssymb}
\usepackage{amsmath}
\usepackage{graphicx}
\usepackage{amsbsy}

\begin{document}

\title{Linear scaling electronic structure calculations and accurate sampling with noisy forces}
\author{Florian R. Krajewski}
\affiliation{Computational Science, Department of Chemistry and Applied Biosciences, ETH
Zurich, USI Campus, Via Giuseppe Buffi 13, CH-6900 Lugano, Switzerland}
\author{Michele Parrinello}
\affiliation{Computational Science, Department of Chemistry and Applied Biosciences, ETH
Zurich, USI Campus, Via Giuseppe Buffi 13, CH-6900 Lugano, Switzerland}
\date{\today}

\begin{abstract}
Numerical simulations based on electronic structure calculations are finding
ever growing applications in many areas of physics. A major limiting factor
is however the cubic scaling of the algorithms used. Building on previous
work [F. R. Krajewski and M. Parrinello, {Phys.Rev. B} \textbf{71}, 233105
(2005)] we introduce a novel statistical method for evaluating the
inter-atomic forces which scales linearly with system size and is applicable
also to metals.
The method is based on exact decomposition of the
fermionic determinant and on a mapping onto a field theoretical expression.
We solve exactly the problem of sampling the Boltzmann distribution with noisy
forces. This novel approach can be used in such diverse fields as quantum
chromodynamics, quantum Monte Carlo or colloidal physics. 
\end{abstract}

\pacs{71.15.-m, 31.15.-p}
\keywords{electronic structure, linear scaling, tight binding, Langevin
equation}
\maketitle

%\preprint{}




%\thanks{}
% \altaffiliation{}






Atomistic simulations in which the interactions are computed on the fly from
electronic structure calculation play an important role in modern science
and have proven their relevance in many fields. However their computational
cost is a severe limitation. In particular simulating large systems has
proven challenging due to the cubic dependence of the computation time on
the number of electrons. This has long since been recognized\cite{yang91,gall92} 
and a number of algorithms have been suggested that in
principle lead to linear 
scaling\cite{goe99,Pay03,Scu99,Kohn96,Gill95,Van93,goedMet,HG}. 
Most are based on the
possibility in semiconductors or insulators of localizing the electronic
orbitals. Linear scaling is then achieved by neglecting interactions between
faraway atoms. This approach however suffers from poor convergence and leads
to errors that are not easy to control. In metals the wavefunctions cannot
be localized and only a handful of methods have been suggested\cite{goedMet,HG}. 
All in all it can be stated that in spite of considerable
progress performing linear scaling ab-initio simulations is still very
challenging.

Very recently we have proposed a new algorithm which scales linearly
in all physical dimensions for semiconductors and metals alike\cite{ON05}. 
Here we  reformulate the algorithm  as a field theory. Sampling the
resulting action is done stochastically. 
We show that, in spite of the
statistical noise present in the evaluation of the forces,
exact sampling can be performed. 
Our way of solving this problem is general and can also solve problems 
of similar nature that are encountered 
in quantum chromodynamics\cite{latt,kennedy}, 
quantum Monte Carlo\cite{pierl04} and colloidal physics\cite{colloids}, 
where also the interaction can only be determined stochastically.

Let us start with the generic expression for the total energy in theories
that can be formulated in an effective single particle form: 
\begin{equation}
E=2\sum_{i=1}^{N}\epsilon _{i}+V_{r}\;.
\end{equation}
The first term is the so-called band structure term given by the sum of the
lowest $N$ doubly occupied eigenvalues $\epsilon _{i}$ of a Hamiltonian $
\mathbf{H}$. For instance in density functional theory $\mathbf{H}$ is the
Kohn and Sham Hamiltonian and $V_{r}$ corrects for double counting and
accounts for the direct nuclear nuclear interaction, 
while in tight binding
and other semi-empirical approaches $\mathbf{H}$ is a Hamiltonian 
which depends parametrically on the atomic positions 
and $V_{r}$ is a pairwise repulsive energy. In either case $V_{r}
$ can be calculated in $\mathcal{O}\left( N\right) $ operations, while the
calculation of the band structure term has an apparent $\mathcal{O}\left(
N^{3}\right)$ complexity and has been the limiting factor that has so far
prevented simulating very large systems.

Following ref.~\cite{alavi92,alavi94} we write the band structure term as
the low-temperature limit of the grand canonical potential for independent
fermions: 
\begin{equation}
\Omega =-\frac{2}{\beta }\ln \det \left( 1+e^{\beta (\mu -\mathbf{H}
)}\right) \;.  \label{eq:gkpot}
\end{equation}
In Eq.~(\ref{eq:gkpot}) $\mu $ is the electron chemical potential and it
easy to see that $\lim_{\beta \rightarrow \infty
}\Omega=2\sum_{i=1}^{N}\epsilon _{i}-\mu N_{e}$, where $N_{e}=2N$ is the
total number of electrons. We now factorize the operator in Eq.~(\ref{eq:gkpot}) 
as: 
\begin{equation}
1+e^{-\beta \left( \mu -\mathbf{H}\right) }=\prod_{l=1}^{P/2}\left( \mathbf{M%
}_{l}^{\dagger }\mathbf{M}_{l}\right)  \label{eq:fac}
\end{equation}
where $P$ is an even integer and 
$\mathbf{M}_{l}=1+e^{i\pi (2l-1)/P}e^{\frac{\beta }{P}\left( \mu -\mathbf{H}\right) }$. 
Here we depart from ref.~\cite{ON05} 
and since $\mathbf{M}_{l}^{\dagger }\mathbf{M}_{l}$ is a positive
definite operator we can follow the practice well known in lattice gauge
simulations\cite{latt} of writing its inverse determinant as an integral
over a field $\mathbf{\Phi }_{l}$ that has the dimension of the full Hilbert
space in the form\cite{fn}: 
\begin{equation}
\mathrm{det}\left( \mathbf{M}_{l}^{\dagger }\mathbf{M}_{l}\right) ^{-\frac{1%
}{2}}=\frac{\int \mathcal{D}[\mathbf{\Phi }_{l}]e^{-\mathbf{\Phi }%
_{l}^{\dagger }\mathbf{M}_{l}^{\dagger }\mathbf{M}_{l}\mathbf{\Phi }_{l}}}{%
\int \mathcal{D}[\mathbf{\Phi }_{l}]e^{-\mathbf{\Phi }_{l}^{\dagger }\mathbf{%
\Phi }_{l}}}\;.  \label{eq:fidet}
\end{equation}
Inserting the relation (\ref{eq:fac}) into Eq.~(\ref{eq:gkpot}) after having
used Eq.~(\ref{eq:fidet}) we end up with the following expression for the
grand canonical potential: 
\begin{equation}
\Omega =\frac{4}{\beta } \sum_{l=1}^{P/2}\ln \int \mathcal{D}[\mathbf{\Phi }%
_{l}] \; e^{-\mathbf{\Phi }_{l}^{\dagger }\mathbf{M}_{l}^{\dagger } \mathbf{M%
}_{l}\mathbf{\Phi }_{l}}+\mathrm{const.}  \label{eq:ourO}
\end{equation}
which is the promised field theoretical formulation. The quantities of
physical interest like energy or force can all be calculated as derivatives
of $\Omega $ relative to an appropriate external parameter. For instance 
$N_{e}=-\frac{\partial }{\partial \mu }\Omega $ , and assuming that $\beta
^{-1}$is so low that temperature effects on the electrons can be neglected 
$E^{\mathrm{band}}=  2\sum_{i=1}^{N}\epsilon _{i}=\frac{\partial }{\partial
\beta }\left( \beta \Omega \right)  +\mu N_{e}$ while the contribution to
force on particle $I$ at position $\mathbf{R}_{I}$ coming from the band term
is given by 
$\mathbf{F}_{I}^{\mathrm{band}}=- \boldsymbol{\nabla}_{ \mathbf{R}_{I} } \Omega $. 
In taking the derivatives the constant in Eq.~(\ref{eq:ourO}) 
vanishes and one is left with calculating expressions of the
type: 
\begin{widetext}
\begin{equation}\label{eq:Oder}
\frac{\partial \Omega}{\partial \lambda}= 
\frac{4}{\beta} \sum_{l=1}^{P/2} 
\frac{ 
\int \mathcal{D}[\mathbf{\Phi}_l] 
\; 
\left[ \mathbf{\Phi}_l^\dagger 
\left( 
      \frac{\partial \mathbf{M}_l^\dagger}
           {\partial \lambda} 
      \mathbf{M}_l 
      + \mathbf{M}_l^\dagger
      \frac{\partial\mathbf{M}_l}
       {\partial\lambda} 
\right) 
\mathbf{\Phi}_l  \right]
\; 
e^{- \mathbf{\Phi}_l^\dagger \mathbf{M}_l^\dagger \mathbf{M}_l \mathbf{\Phi}_l} 
}{ 
\int \mathcal{D}[\mathbf{\Phi}_l] \;
e^{- \mathbf{\Phi}_l^\dagger \mathbf{M}_l^\dagger \mathbf{M}_l \mathbf{\Phi}_l} } \;. 
\end{equation}
\end{widetext}
Thus all relevant properties can be evaluated by sampling the $P/2$
distributions 
$e^{-\mathbf{\Phi }_{l}^{\dagger }\mathbf{M}_{l}^{\dagger}\mathbf{M}_{l}\mathbf{\Phi }_{l}}$. 

So far no approximation has been made
and no computational advantage has been gained either. In order to make
further progress we must take advantage of the fact that in $\mathbf{M}_{l}$
the operator $e^{\frac{\beta }{P}\left( \mu -\mathbf{H}\right) }$ appears
and that $P$ can be taken to be sufficiently large for suitable
approximations to the exponential operator to be accurate. In ref.~\cite%
{ON05} we used as basis set a grid in real space and a Trotter decomposition
was the natural approximation to use. Here we will apply our method to a
tight binding Hamiltonian and simply use a high-temperature expansion: 
\begin{equation}
\mathbf{M}_{l}=1+e^{\pi (2l-1)/P}\left[ 1+\frac{\beta }{P}\left( \mu - 
\mathbf{H}\right) \right] +\mathcal{O}\left( \left( \frac{\beta }{P}\right)
^{2}\right) \;.  \label{eq:MApprox}
\end{equation}
In such a manner the operator $\mathbf{M}_{l}$ has the same sparsity of $%
\mathbf{H}$, a fact which will eventually lead to linear scaling. We have
also considered the possibility of using higher order expressions for the
exponential operator in Eq.~(\ref{eq:MApprox}) \cite{goedMet}. This leads to
a smaller $P$ value at the cost of a less sparse $M_l$. Since eventually we
want to exploit parallelism as much as possible we have preferred to use the
expression~(\ref{eq:MApprox}). Note that no assumption has been made on the
energy spectrum or on the local character of the wavefunctions and therefore
our method will be valid both for metals and non-metals. It simplifies the
calculation of the properties of the system if in Eq.~(\ref{eq:Oder}) we use
the expression 
\begin{equation}
\frac{\partial \mathbf{M}_{l}}{\partial \lambda }=\frac{1}{2P}\left\{ \left( 
\mathbf{M}_{l}-1\right) \mathbf{O}_{\lambda }+\mathbf{O}_{\lambda }\left( 
\mathbf{M}_{l}-1\right) \right\} +\mathcal{O}\left( \frac{1}{P^{3}}\right)
\;.  \label{eq:MDer}
\end{equation}
which has an accuracy compatible with Eq.~(\ref{eq:MApprox}). In Eq.~(\ref{eq:MDer}) 
$\mathbf{O} _{\lambda }$ can be $\beta $, $(\mu -\mathbf{H})$, or 
$-\beta \boldsymbol{\nabla}_{\mathbf{R}_{I}} \mathbf{H} $ for $\lambda =\mu $, 
$\beta $ or $\mathbf{R}_{I}$ respectively. A standard approach to sampling 
$e^{-\mathbf{\Phi }_{l}^{\dagger }\mathbf{M}_{l}^{\dagger }\mathbf{M}_{l}\mathbf{\Phi }_{l}}$ 
is to draw a sequence of normal distributed random
numbers $\mathbf{\Psi }_{l}$ and compute $\mathbf{\Phi }_{l}$ solving the
equations $\mathbf{M}_{l}\mathbf{\Phi }_{l}=\mathbf{\Psi }_{l}$. Since $\mathbf{M}_{l}$ 
is sparse, this equation can be solved in $\mathcal{O}\left(
N\right) $ operations using for instance a biconjugated gradient method\cite{num_rec}. 
It can be easily shown that this approach is equivalent to the
stochastic inversion method advocated in ref.~\cite{ON05}. Here we have
decided instead to sample 
$e^{-\mathbf{\Phi }_{l}^{\dagger }\mathbf{M}_{l}^{\dagger }\mathbf{M}_{l}\mathbf{\Phi }_{l}}$ 
using a Langevin dynamics. 
\begin{equation}
m_{l}\ddot{\mathbf{\Phi }}_{l}=-\mathbf{M}_{l}^{\dagger }\mathbf{M}_{l} 
\mathbf{\Phi }_{l}-\gamma _{e}\dot{\mathbf{\Phi }}_{l}+\boldsymbol{\xi }_{l}
\end{equation}
where the components $\alpha $ of the white random noise vector $\mathbf{\xi 
}_{l}$ obey the relation 
\begin{equation}
\left\langle \mathbf{\xi }_{l}^{\alpha }(0)\mathbf{\xi }_{l}^{\alpha
}(t)\right\rangle =2 m_{l}\gamma _{e}\delta (t)\;.
\end{equation}
In this way we circumvent the need to invert the matrix $\mathbf{M}_{l}$.
This solves the problem that for a tight binding Hamiltonian as opposed to
the Hamiltonian used in ref.~\cite{ON05} we were not able to find good
preconditioners. Furthermore since different $\mathbf{M}_{l}$'s have
different eigenvalue spectra one can choose the $m_{l}$ so as to achieve the
optimal sampling speed in each $l$ channel. This problem is particularly
serious for metals where $\mathbf{M}_{l\thickapprox \frac{P}{2}}$ can have
eigenvalues close to zero, which leads to a much slower sampling 
speed\cite{ON05}. 
Finally since we will use a Langevin sampling also for the ions it
is pleasing to use the same sampling methodology for electronic and ionic
degrees of freedom.

Inevitably the interatomic forces that are calculated by this procedure will
be affected by a statistical error. This will prevent us from using these
forces to perform energy conserving MD calculations. However we will show
that sampling of the Boltzman distribution is still possible. Similarly to
what was done for the electronic degrees of freedom we sample the ionic
configurations with a Langevin equation: 
\begin{equation}
M\ddot{\mathbf{R}}_{I}= \mathbf{F}_{I}-\gamma_I \dot{\mathbf{R}}_{I}+\mathbf{
\Xi }_{I}  \label{eq:leq}
\end{equation}
where the random force obeys the relations 
\begin{equation}
\left\langle \mathbf{\Xi }_{I}(0)\mathbf{\Xi }_{I}(t)\right\rangle = 6k_{
\mathrm{B}}T M \gamma_I \delta (t)
\end{equation}
and 
\begin{equation}
\left\langle \mathbf{F}_{I}(0)\mathbf{\Xi }_{I}(t)\right\rangle =0\;.
\end{equation}
From the electrons' Langevin dynamics we do not get the exact forces 
$\mathbf{F}_{\mathrm{I}}$ but an approximation 
$\mathbf{F}_{\mathrm{I}}^{L}=\mathbf{F}_{\mathrm{I}}+\mathbf{\Xi }_{\mathrm{I}}^{L}$ 
that is affected by
a statistical error $\mathbf{\Xi }_{I}^{L}$ and therefore there is in
principle no guarantee that correct Boltzman averages are obtained from the
solution of Eq.~(\ref{eq:leq}). But let us assume, and we will show later in
a realistic case that this assumption is indeed justified, that also 
$\mathbf{\Xi }_{I}^{L}$ is a white noise obeying 
\begin{eqnarray}
\left\langle \mathbf{\Xi }_{I}^{L}(0)\mathbf{\Xi }_{I}^{L}(t)\right\rangle
&\cong & 6k_{\mathrm{B}} T M \gamma ^{L}_I \delta (t)\quad \mathrm{and} 
\notag  \label{eq:ncond} \\
\left\langle \mathbf{F}_{I}(0)\mathbf{\Xi }_{I}^{L}(t)\right\rangle &\cong&
0\;.
\end{eqnarray}
In this case the noise $\mathbf{\Xi }_{\mathrm{I}}^{L}$ simply adds to $%
\mathbf{\Xi }_{\mathrm{I}}$ and if we modify Eq.~(\ref{eq:leq}) so as to
read 
\begin{equation}
M \ddot{\mathbf{R}}_{I} =\mathbf{F}_{I}-\left( \gamma_I +
\gamma_I^{L}\right) \dot{\mathbf{R}}_{I} +\mathbf{\Xi }_{I}+\mathbf{\Xi }
_{I}^{L}
\end{equation}
we recover a Langevin equation whose trajectories can still be used to
obtain a Boltzman sampling. 

At first sight it would appear that we are
defeating our object since in general we know 
$\mathbf{F}_{I}+{\mathbf{\Xi }}_{I}^{L}$ but not each term individually. 
However we can determine 
$\gamma^{L}_I$ 
by varying it until the equipartition theorem $\left\langle 
\frac{1}{2}M\dot{\mathbf{R}}_{I}^{2}\right\rangle =\frac{3}{2}k_{\mathbf{B}}T
$ is satisfied. With this choice the sampling will be correct and noisy
forces can be used in the sampling. It is important to note that after the
correct value for $\gamma^L_I $ is determined it has to be kept constant
during the simulation. With this procedure one can exactly calculate static
observables within the framework of Langevin dynamics without knowing the
exact force. This is at variance with ref.~\cite{cep} where explicit
assumptions on the noise have to be made.

It remains now to show that the assumptions made above are correct and that
the resulting scheme is accurate. As a test system we have chosen silicon in
the diamond and in the liquid phase. We use the empirical tight binding
interaction for silicon described in \cite{TBham}. 
We checked that a value
of $P=200$ is sufficient for the approximation 
of Eq.~(\ref{eq:MApprox}) to be valid. The Langevin
dynamics parameters used were $\delta t_e=1$, $\gamma _e\delta t_e=1/20$
where $\delta t_e$ is the discretized integration time step and the
algorithm of Ricci and Ciccotti \cite{LD} has been used throughout. The
masses $m_{l}$ are adjusted such that the average force fluctuations are 
$\left(  \mathbf{M}_{l}^{\dagger }\mathbf{M}_{l}\mathbf{\Phi }_{l}  \right)^2
\frac{ \delta t_e^{2} }{2m_{l}}  <0.025  $. This adjustment of the masses is
a crucial ingredient to obtain a good performance since if the masses were
set to be equal for all values for $l$ the fastest time scales of the field 
$\Phi _{P/2}$ would be more than one order of magnitude larger than that of 
$\Phi _{1}$. The integration time step for the ion dynamics is 
$\delta t=1\,\mathrm{fs}$. After each ionic displacement we let the $\mathbf{\Phi}_l$
evolve under the action of the new $\mathbf{M}^\dagger_l \mathbf{M}_l$ until
the distribution is equilibrated. The time needed for the equilibration is
problem dependent and can be measured by looking at the correlation function 
$\left< \mathbf{\Phi}_l(0) \mathbf{\Phi}_l(t) \right>$. In the present case
we make the rather conservative choice of running the electronic Langevin
equation for 100 time steps. The $\mathbf{\Phi_l}$ thus obtained are used to
calculate the ionic forces for the next integration step. The chemical
potential is continuously adjusted such that the number of electrons
fluctuates around the desired value.

\begin{figure}[t]
\begin{center}
\includegraphics[width=8.5cm]{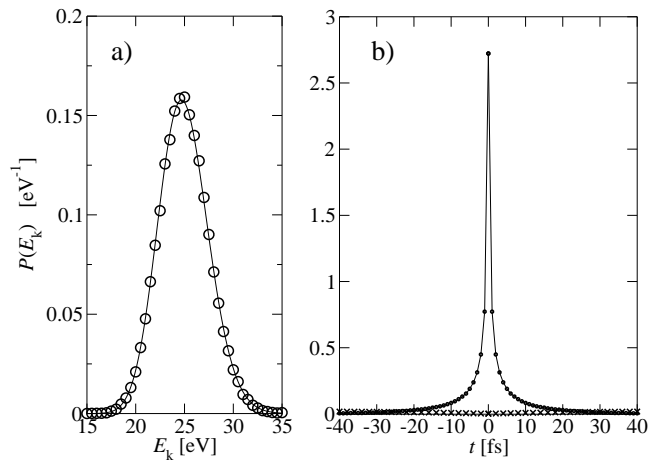}
\end{center}
\caption{ a) Distribution of the kinetic energy for a simulation of liquid
silicon with 64 atoms at 3000K (circles). Maxwell distribution (line). b)
Autocorrelation function of the noise 
$\left\langle \mathbf{\Xi }_{I}^{L}(0)\mathbf{\Xi }_{I}^{L}(t)\right\rangle / 
\left\langle \left( \mathbf{F}_{I}^{\mathrm{band}}(0) \right)^2 \right\rangle $ 
(line). Cross-correlation function of the noise and the exact force 
$\left\langle \mathbf{F}_{I}(0)
\mathbf{\Xi }_{I}^{L}(t)\right\rangle / \left\langle \left( \mathbf{F}_{I}^{
\mathrm{band}}(0) \right)^2 \right\rangle$ (crosses). The correlation
functions are normalized to the exact ionic force stemming from the band
energy. All results are calculated for a system 64 atoms of liquid silicon
at 3000K. }
\label{fig:noise}
\end{figure}

\begin{figure}[t]
\begin{center}
\includegraphics[width=8.5cm]{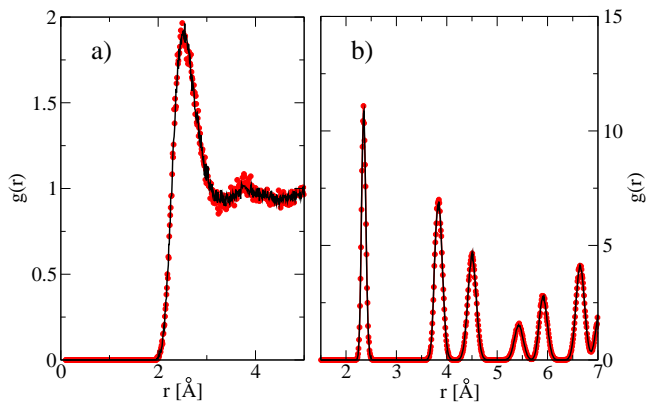}
\end{center}
\caption{ (color) Pair correlation functions for liquid silicon (3000K) (a)
and crystalline silicon in the diamond phase (300K) (b). The red dots show
the results from our new method and the lines are calculated using standard
diagonalisation of the tight binding Hamiltonian. }
\label{fig:pairCorr}
\end{figure}
We first consider the case of 64 Si atoms in a periodically repeated cell of
length 10.86 \AA\ at a temperature of $3000\;K$ where the system is
metallic. Using the procedure described above we find that if we take $%
\gamma_I =\frac{1}{30}\,\mathrm{fs}^{-1}$ we need to add a correction due to
the noise in the forces $\gamma ^{L}_I=\frac{1}{379}\,\mathrm{fs}^{-1}$
which fixes the average kinetic energy of the particles at the desired
value. In Fig.~\ref{fig:noise}~a) it is also seen that not only is the
average energy correct but also that its fluctuations follow Maxwell
distribution. In this small system it is possible to calculate the correct
forces on the ions and check the statistical properties of 
$\mathbf{\Xi }_{I}^{L}$. 
It is seen from Fig.~\ref{fig:noise}~b) that Eq.~(\ref{eq:ncond})
is satisfied to a very good approximation and that the correlation between
the exact forces and the noise is minimal. Furthermore 
$\left\langle \mathbf{\Xi }^{L}_I(0)\mathbf{\Xi }^{L}_I(t)\right\rangle $ 
is strongly localized in
time and its behavior is well fitted by two exponentials whose decay is much
faster than the fastest ionic motion time scale. Therefore its net effect is
similar to that of a delta function and in fact if we approximate 
$\left\langle \mathbf{\Xi }^{L}_I(0)\mathbf{\Xi }^{L}_I(t)\right\rangle $ as
a delta function whose strength is given by its integral we find an estimate
for $\gamma^{L}_I\thickapprox \frac{1}{345} \,\mathrm{fs}^{-1}$ in good
agreement with the empirical determination. Similar analysis conducted on
semiconducting Si leads to a similar conclusion, namely that the effect of
the noise in the calculation of the forces can be simply accounted for by
adding the damping coefficient $\gamma^L_I$. In the solid phase we find 
$\gamma^L_I = \frac{1}{502} \,\mathrm{fs}^{-1} $ which is similar to the
result for the liquid. Therefore we expect that, with some adjustment phase
transitions can be studied with our method, at least in this case.

In Fig.~\ref{fig:pairCorr} we compare the pair correlation functions 
$g\left( r\right) $ calculated with noisy forces and those evaluated with a
standard approach. It is seen that the agreement is excellent and that the
use of noisy forces does not degrade the quality of the simulation.

There are two issues that we would like to discuss briefly. One is the
break-even point between standard calculations and the present linear
scaling method. This is not easy to determine since it depends on the
accuracy required, which in our case is related to the number $P$ used in
the decomposition of the fermion determinant Eq.~(\ref{eq:fac}). For the
case of Si we estimate on a single processor the crossing point to be at
about 500 atoms. In other cases larger values of $P$ might be necessary,
thus shifting the crossing point to larger systems. However our algorithm is
trivially parallelizable and therefore we expect the situation to be much
more favorable in terms of wall clock time on massive parallel platforms.
Our code scales essentially linearly with $P$ while the diagonalizion codes,
although highly developed, are not amenable to similarly efficient
parallelization. For instance Shellman et al.\cite{voth} have estimated that
parallelization of a tight binding simulation on 1000 atoms loses efficiency
when distributed over more than 8 processors. In addition we can gain even
more if we organize our calculation hierarchically and distribute the most
expensive part of the calculation, the matrix multiplication 
$\mathbf{M}_{l}^{\dagger }\mathbf{M}_{l}\mathbf{\Phi }_{l}$, to several processors.

Finally we stress once again that the interest in our
work transcends the field of linear scaling algorithms and offers an
alternative to other MC methods with noisy estimators\cite{cep,kennedy}.

We thank A. Laio for his critical reading of the manuscript.


\begin{thebibliography}{99}
\bibitem{fn} An elegant expression for the partition function of the total
system (electrons + ions) can be obtained if one uses the alternate
expression $ \mathrm{det}\left( \mathbf{M}_{l}^{\dagger }\mathbf{M}%
_{l}\right) ^{\frac{1}{ 2}}=\frac{\int \mathcal{D}[\mathbf{\Phi }_{l}]e^{-%
\mathbf{\Phi } _{l}^{\dagger }\left( \mathbf{M}_{l}^{\dagger }\mathbf{M}%
_{l}\right) ^{-1} \mathbf{\Phi }_{l}}}{\int \mathcal{D}[\mathbf{\Phi }%
_{l}]e^{-\mathbf{\Phi } _{l}^{\dagger }\mathbf{\Phi }_{l}}}\;, $ leading to $
Z \propto \int dR_{I} e^{-\beta V_{r}} \int \prod_{l=1}^{P}\mathcal{D}[%
\mathbf{\Phi }_{l}] e^{ -\sum_{l=1}^{P}\mathbf{\Phi }_{l}^{\dagger }\left( 
\mathbf{M}_{l}^{\dagger }\mathbf{M}_{l}\right) ^{-1}\mathbf{\Phi }_{l} }\;. $
Unfortunately this led to an algorithm which was less efficient than what is
described here. This will be discussed in detail elsewhere.
\end{thebibliography}
\end{document}